\documentclass[preprint
 amsmath,amssymb,
 aps, physrev,
]{revtex4-2}
\usepackage{graphicx}
\usepackage{dcolumn}
\usepackage{bm}
\usepackage{float}
\usepackage{color}
\usepackage{soul}
\usepackage{adjustbox}
\usepackage{comment}
\usepackage{natbib}
\usepackage{amsmath}
\usepackage[mathlines]{lineno}

\begin{document}

\title{From point patterns to networks: to what extent does the Delaunay triangulation reproduce key spatial and density information?}

\author{Eli Newby}
\affiliation{Department of Physics, Pennsylvania State University, University Park, PA 16802}
\affiliation{Department of Cardiovascular and Metabolic Sciences, Lerner Research Institute, Cleveland Clinic, Cleveland, OH 44195}
\email{newbye@ccf.org}

\author{Wenlong Shi}
\affiliation{Materials Science and Engineering, Arizona State University, Tempe, AZ 85287}

\author{Yang Jiao}
\affiliation{Materials Science and Engineering, Arizona State University, Tempe, AZ 85287}

\author{Salvatore Torquato}
\affiliation{Department of Chemistry, Princeton University, Princeton, New Jersey 08544, USA}
\affiliation{Department of Physics,  Princeton University, Princeton, New Jersey 08544, USA}
\affiliation{Princeton Institute  of Materials, Princeton University, Princeton, New Jersey 08544, USA}
\affiliation{Program in Applied and Computational Mathematics, Princeton University, Princeton, New Jersey 08544, USA}

\author{R\'eka Albert}
\affiliation{Department of Physics, Pennsylvania State University, University Park, PA 16802}
\affiliation{Department of Biology, Pennsylvania State University, University Park, PA 16802}
\email{rza1@psu.edu}
\date{\today}

\begin{abstract}{
To be used as an analysis tool, it is important that a spatial network’s construction algorithm reproduces the structural properties of the original physical embedding. One method for converting a two-dimensional (2D) point pattern into a spatial network is the Delaunay triangulation. Here, we apply the Delaunay triangulation to seven different types of 2D point patterns, including hyperuniform systems. The latter are characterized by completely suppressed normalized infinite-wavelength density fluctuations. We demonstrate that the quartile coefficients of dispersion of multiple centrality measures are capable of rank-ordering hyperuniform and nonhyperuniform systems independently, but they cannot distinguish a system that is nearly hyperuniform from hyperuniform systems. Thus, in each system, we investigate the local densities of the point pattern $\rho_P(\mathbf{r}_i;\ell)$ and of the network $\rho_G(n_i;\ell)$. We reveal that there is a strong correlation between $\rho_P(\mathbf{r}_i;\ell)$ and $\rho_G(n_i;\ell)$ in nonhyperuniform systems, but there is no such correlation in hyperuniform systems. When calculating the pair-correlation function and local density covariance function on the point pattern and network, the point pattern and network functions are similar only in nonhyperuniform systems. In hyperuniform systems, the triangulation has a positive covariance of local network densities in pairs of nodes that are close together; such covariance is not present in the point patterns. Thus, we demonstrate that the Delaunay triangulation accurately captures the density fluctuations of the underlying point pattern only when the point pattern possesses a positive correlation between $\rho_P(\mathbf{r}_i;\ell)$ for points that are close together. Such positive correlation is seen in most real-world systems, so the Delaunay triangulation is generally an effective tool for building a spatial network from a 2D point pattern, but there are situations (i.e., disordered hyperuniform systems) where we caution that the Delaunay triangulation would not be effective at capturing the underlying physical embedding.}
{Spatial Networks, Hyperuniformity, Delaunay Triangulation, Graph Theory, Complex Networks}
\end{abstract}
\maketitle

\section{\label{sec:intro}Introduction}
When the elements of an interacting system are embedded in physical space, one can represent the system as a spatial network \cite{Barthelemy2011,Daqing2011,Barthelemy2018}. Spatial networks have helped model and analyze a wide range of real-world systems, including power grids, transportation, the internet, granular media, and the brain \cite{Yook2002,Bullmore2009,Farahani2013,Papadopoulos2018,Duchemin2023,Penrose2003,Kurant2006, Barrat2004,Jiao2012}. An area of interest in spatial network analysis is how to build a network that reproduces properties of the physical embedding. Here, we focus on 2D point patterns (i.e., a simple physically embedded system) composed of $N$ points $\bf{R} = \{\mathbf{r}_1, \mathbf{r}_2, \dots \mathbf{r}_N\}$ where $\mathbf{r}_i$ is the coordinate of point $i$. During the conversion of a 2D point pattern into a network, we define the nodes of the network as the points of the point pattern, and aim to define edges in such a way as to capture the properties of the point pattern such as the local density surrounding each point. That is, for a spatial network $\mathcal{G}(\mathcal{N},\mathcal{E})$ built from the point pattern $\bf{R}$, we are interested in how each node $n_i \in \mathcal{N}$ compares to the point at position $\mathbf{r}_i$ that originates it. Multiple ways of converting a 2D point pattern into a network have been studied \cite{Barthelemy2011, Barthelemy2018,Waxman1988,Dettmann2016,Duchemin2023,Rozenfeld2002,Andrade2005}. For example, geometric graphs are spatial networks formed by adding an edge between every pair of points whose distance is smaller than a specified connection distance. The mathematical and statistical properties of geometric graphs built from point patterns distributed uniformly at random have been substantially researched, and these graphs have been used in a variety of modeling contexts \cite{Dall2002, Penrose2003, Kenniche2010, Dettmann2016, Duchemin2023}.

Another technique for obtaining a network from a 2D point pattern is constructing the Delaunay triangulation of the point pattern \cite{Delaunay1934, Lee1980, Su1997}. The Delaunay triangulation connects the points into a planar, triangular network such that the circumcircles of every triangle do not contain any other points. This triangulation algorithm is typically preferred because the angles of the triangles in the resulting network are maximized (i.e., the construction avoids the creation of sliver triangles). In contrast to geometric graphs, the Delaunay triangulation is parameterless and creates a planar network from the point pattern. As a result, this construction algorithm better captures the interactions between a point and its immediate neighbors in space, which makes this method preferable in certain modeling situations of 2D point patterns.

Previous work investigating the structural properties of the Delaunay triangulation has provided insight into its theoretical limitations \cite{Lee1986, Chew1987, Musin2003, Aurenhammer2013}. One open problem is identifying the maximum possible tortuosity (i.e., the stretch factor) of the triangulation. In graph theory, this open problem is also known as identifying the spanning ratio $t$ of the Delaunay triangulation (i.e., for a set of points defined in space, a t-spanner is a network in which every path through the system has tortuosity $\leq t$) \cite{Xia2011, Xia2013, Perkovic2022, Tan2022, Bonichon2015, Bonichon2023}. The tortuosity is the ratio of the network distance (i.e., the geodesic distance) and the Euclidean distance between a pair of points. Previous research has aimed to put bounds on the Delaunay triangulation’s maximum possible tortuosity. Over the past few years, the upper bound on the maximum tortuosity of the Delaunay triangulation has steadily decreased; currently it is shown that for a convex point pattern the upper bound is 1.84 \cite{Tan2022}. The lower bound of the maximum tortuosity has also increased from its original value of $\pi/2$ to its current value of 1.5932 \cite{Xia2011}. Because there is still a gap between these two values, the question of the Delaunay triangulation’s maximum tortuosity remains an open problem for computational geometry. 

This previous work is important for better understanding the worst-case scenarios of the Delaunay triangulation’s distortion of distance, but it does not address how effective the Delaunay triangulation is in typical situations. Here, we investigate the properties of the Delaunay triangulation when constructed for various 2D point patterns that reflect real systems. We determine how the underlying structure of each point pattern affects the structures of the triangulations. We specifically compare the Delaunay triangulations constructed on two classes of systems: hyperuniform point patterns and nonhyperuniform point patterns. Hyperuniform systems are a state of matter characterized by the suppression of long-range density fluctuations \cite{Torquato2003, Torquato2016, Torquato2018, Zheng2020, Chen2023, Milosevic2019, Tang2022, Liu2024, chen2025anomalous}. They consist of ordered hyperuniform systems, such as crystals, and the exotic state of disordered hyperuniform systems, which possess properties of crystals (e.g., suppression of infinite-wavelength density fluctuations) and also properties of liquids (e.g. isotropy). Because of these dual properties, disordered hyperuniform systems are thought to have a “hidden order” \cite{Klatt2019, Zheng2021}. We are interested in the degree to which spatial network methods reproduce the complexity of these exotic states of matter, if they are able to capture the hidden order of disordered hyperuniform systems, and how the results of spatial network methods applied to disordered hyperuniform systems compare to the results of their application to better-studied (e.g., Poisson) systems.

\begin{figure*}[htpb]
    \includegraphics[width=\textwidth]{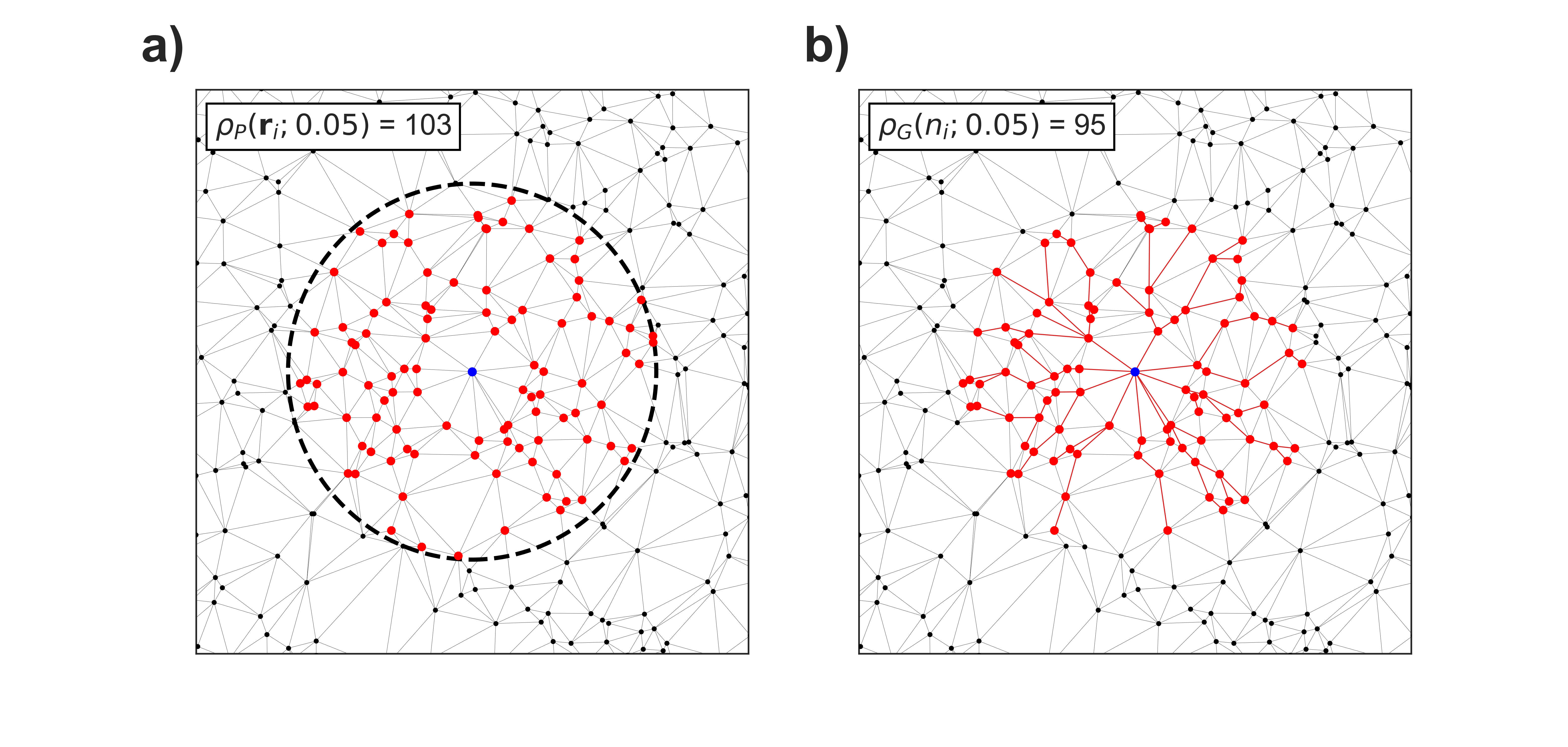}
    \caption{a) The local point pattern density $\rho_P(\mathbf{r}_i;\ell)$ of a point at position $\mathbf{r}_i$ (indicated in blue) for $\ell = 0.05$ is the number of points (indicated in red) within a circle of radius $\ell$ (indicated by the dotted circle). b) The local network density $\rho_G(n_i;\ell)$ of a node $n_i$ corresponding to the point at position $\mathbf{r}_i$ (indicated in blue) for $\ell = 0.05$ is the number of nodes within a distance of $\ell$ along the edges in the network (indicated in red). The inset in each panel indicates the local density, and the point pattern has a greater local density than the network. This is because distance in the network is restricted to the edges of the system, so $\rho_P(\mathbf{r}_i;\ell) \geq \rho_G(n_i;\ell)$.}
    \label{Fig:LocalDensity}
\end{figure*}

We show that various centrality measures are able to correctly sort triangulations based on their level of order, but they are unable to completely differentiate between structurally similar hyperuniform and nonhyperuniform systems. To uncover the reasons for this imperfect sensitivity, we investigate whether the Delaunay triangulation distorts the density of these systems and, if there is distortion, what mechanisms cause this distortion. We perform this density investigation by studying the local point pattern density $\rho_P(\mathbf{r}_i;\ell)$ of every point at position $\mathbf{r}_i$ and local network density $\rho_G(n_i;\ell)$ of every node $n_i$ for each system. The local density $\rho_P(\mathbf{r}_i;\ell)$ of a point $\mathbf{r}_i$ is the number of points within Euclidean distance $\ell$ from the point (i.e., the number of points within a circle of radius $\ell$ centered at $\mathbf{r}_i$), and the local density $\rho_G(n_i;\ell)$ of a node $n_i$ is the number of points within a network distance $\ell$ (i.e., a traversal along the edges that is smaller than $\ell$) (see Figure \ref{Fig:LocalDensity} and Key Concepts for more details). Because distance is only measured along the edges of the network, distances between points in the network differ from the distances in the point pattern. As a result, local densities are not equivalent between the points and their respective nodes $\rho_P(\mathbf{r}_i;\ell) \neq \rho_G(n_i;\ell)$. We are interested in understanding how these different measures of distance distort the density, which will affect the triangulation’s reproduction of structural properties. 

Our analysis of local densities indicates distinct differences between nonhyperuniform and hyperuniform systems. We find that in nonhyperuniform systems $\rho_P(\mathbf{r}_i;\ell)$ is highly correlated to $\rho_G(n_i;\ell)$, yet it is uncorrelated in hyperuniform systems. Comparison of the network pair-correlation function to the pair-correlation function of the point pattern indicates that density distortions are slightly larger in hyperuniform systems. We investigate the ability of the Delaunay triangulation to capture the density fluctuations of each system using local density covariance functions $C_\rho(r;\ell)$ (see Key Concepts). For nonhyperuniform systems, the local density covariance function is similar between the point pattern and the network. On the other hand, we find that, in the point pattern, hyperuniform systems have little to no correlation between $\rho_P(\mathbf{r}_i;\ell)$ for pairs of points at any distance, but, in the network, hyperuniform systems have a positive correlation of $\rho_G(n_i;\ell)$ for pairs of nodes separated by short distances. This indicates that the connected nature of the network prevents the network density from fluctuating as quickly as in the point pattern. Thus, the derived network’s density does not reflect the density of its underlying point pattern. In summary, we show  in this manuscript that the Delaunay triangulation reproduces the point pattern’s density information in nonhyperuniform systems, which reflect more typical real-world systems, but it is ineffective at capturing the unique density properties of disordered hyperuniform systems. Thus, in most applications, the Delaunay triangulation is an effective tool for utilizing a network-based approach to study physically embedded data.

This manuscript is organized in the following manner. In section \ref{sec:methods}, we present key concepts: hyperuiformity, graph definitions, local density, and the various network measures we employ to study the triangulations. In section \ref{sec:Models}, we present the various models that we studied in our investigation. In section \ref{sec:Results}, we present the results of our applications of these network measures to the triangulations. Finally, in section \ref{sec:Discussion}, we interpret our results and discuss the conclusions we draw.

\section{\label{sec:methods}Key Concepts}
\subsection{\label{subsec:Hyperuniformity} Hyperuniformity and Nonhyperuniformity}
Hyperuniformity is a novel state of matter that is currently studied in physical space \cite{Torquato2018, Torquato2003, Zachary2009, Torquato2016}. Hyperuniform many-particle systems are characterized by the suppression of the system’s long-range density fluctuations \cite{Torquato2003, Torquato2018}. For a many-particle system in d-dimensional Euclidean space, we can determine the class of system based on the following variance scaling:
\begin{equation}
\label{equ:Hyperuniform}
\lim_{v(R)\rightarrow\infty}{\frac{\sigma^2_N(R)}{v(R)} = \begin{cases}
0 & Hyperuniform \\
C & Nonhyperuniform \\
\infty & Antihyperuniform
\end{cases}}
\end{equation}

Here, $v(R)$ is the volume of a $d$-dimensional sphere of radius $R$. That is, in hyperuniform systems the variance in the number of particles inside a $d$-dimensional sphere of radius $R$ increases slower than the volume of the hypersphere, in nonhyperuniform systems the number of particles and volume of the hypersphere increase at the same rate, and in anti-hyperuniform systems the number of particles increases faster than the volume of the hypersphere. Here we focus on point patterns defined in 2-dimensional space, so we compare the variance in the number of points that fall within circles of radius $R$ to the areas of these circles. Hyperuniformity can also be defined in terms of the structure factor $S({\bf k})$, which is calculated based on the Fourier transform of the pair-correlation function $g_2({\bf r})$ \cite{Torquato2018}, i.e.,
\begin{equation}
S({\bf k}) = 1+\int_{\mathbb{R}^d}[g_2({\bf r})-1]\exp(-i {\bf k}\cdot{\bf r})d{\bf r}.
\end{equation}
Each system is defined based on $lim_{{\bf k} \rightarrow 0} S({\bf k})$, with hyperuniform systems having $lim_{{\bf k} \rightarrow 0} S({\bf k}) = 0$, nonhyperuniform systems having $lim_{{\bf k} \rightarrow 0} S({\bf k}) = const.$, and antihyperuniform systems having $lim_{{\bf k} \rightarrow 0} S({\bf k}) = \infty$. The hyperuniform systems are categorized into three classes based on the large-$R$ scaling behavior of the number variance \cite{Torquato2018}:
\begin{equation}
\sigma^2(R) \sim \begin{cases}
R^{d-1}, \quad\quad\quad \alpha >1 \qquad &\text{(Class I)}\\
R^{d-1} \ln R, \quad \alpha = 1 \qquad &\text{(Class II)}\\
R^{d-\alpha}, \quad 0 < \alpha < 1\qquad  &\text{(Class III).}
\end{cases}
\end{equation}
For this investigation, all hyperuniform point patterns belong to Class I.



\subsection{\label{subsec:Networks}Graphs and Networks}
A graph, or synonymously, network $\mathcal{G}(\mathcal{N},\mathcal{E})$ is defined by a set of nodes $\mathcal{N}$ of size $N$ and its set of connecting edges $\mathcal{E}$ where an edge indicates a local interaction or relationship between a pair of nodes. Here, we create weighted networks from point patterns, where every edge $E \in \mathcal{E}$ possesses a weight attribute equivalent to the Euclidean distance between the two nodes connected by the edge. This weight allows us to compare the network to the underlying point pattern. We use these weighted networks for our analysis.

\subsection{\label{subsec:LocalDensity}Local Density}
The local density $\rho(\mathbf{r};\ell)$ around the location $\mathbf{r}$ in the point pattern is the number of points within a distance $\ell$ from that location. Since we are interested in comparing the points of the point pattern $\mathbf{r}_i$ to the corresponding nodes in the network $n_i$, we focus on calculating the local density around each point in the network $\rho(\mathbf{r}_i;\ell)$. It can be represented mathematically as:
\begin{equation}
\rho_P(\mathbf{r}_i;\ell) = \sum_{j\in \bf{R}} \Theta(\ell-d(\mathbf{r}_i,\mathbf{r}_j))
\end{equation}
Here, $\Theta(x)$ is the Heaviside step function, and $d(\mathbf{r}_i,\mathbf{r}_j)$ is the distance between point $i$ and point $j$.

The definition of the local network density is similarly:
\begin{equation}
\rho_G(n_i;\ell) = \sum_{j\in \mathcal{N}} \Theta(\ell-d(n_i,n_j))
\end{equation}
However $d(n_i,n_j)$ is slightly different from the point pattern definition of distance. In the point pattern $d(\mathbf{r}_i,\mathbf{r}_j)$ is the Euclidean distance between the two points. In an unweighted network, the network distance between two nodes $d(n_i, n_j)$ is the fewest edges needed to traverse from $n_i$ to $n_j$. In our weighted systems, the distance $d(n_i, n_j)$ is instead the smallest sum of edge weights (i.e., the Euclidean distances between the edges' points) along a path in the network between these two nodes (i.e., the geodesic distance). As a result, the local network density $\rho_G(n_i)$ of every node is slightly distorted because of this different definition of distance.

The differences between the point pattern and network density are presented in Fig. \ref{Fig:LocalDensity}, which shows the local point pattern density $\rho_P(\mathbf{r}_i)$ (a) and local network density $\rho_G(n_i)$ (b) for a point $\mathbf{r}_i$ and its corresponding node $n_i$ (blue point in the panels). The points included in the calculation of local density are highlighted in red in each panel. In the point pattern (a), the points included in the local density calculation are all points within a circle of radius $\ell$. In the network (b), only the paths highlighted in red have a network distance less than $\ell$ from the central node. There are less points within distance $\ell$ in the network than the point pattern.

\subsection{\label{subsec:Centrality}Centrality Metrics}
The nodes of a network can be characterized by a variety of centrality measures, each representing a different flavor of importance. Below, we present the formalization of the seven node centrality measures we calculate on each Delaunay triangulation. Along with investigating the distributions of these seven node centrality measures across all nodes in the network, we also report the distribution of edge weights; this can be considered an edge centrality measure, with shorter edges being more central.

\subsubsection{\label{subsubsec:Strength}Strength}
The strength $T(n_i)$ of a node $n_i$ is a measure of how connected a node is to its surrounding nodes in the network. It is an extension of the degree to weighted systems. It is calculated as 
\begin{equation}
\label{equ:Strength}
    T(n_i) = \sum_{n_j\in E_i} \frac{1}{d(n_i,n_j)}
\end{equation}
where $E_i$ is the set of nodes that are connected with an edge to node $n_i$, and $d(n_i,n_j)$ is the distance between nodes $n_i$ and $n_j$ (i.e., the weight of the edge).

\subsubsection{\label{subsubsec:NodesWithin}Nodes Within Distance $\ell$}
The nodes within distance $\ell$ on the network is equivalent to the local network density of the node, so for a node $n_i$ the number of nodes within distance $\ell$ is $\rho_G(n_i;\ell)$

\subsubsection{\label{subsubsec:Closeness}Closeness Centrality}
The closeness centrality $C(n_i)$ is a centrality measure that measures how close a node $n_i$ is to the rest of the network. It is calculated by finding the sum of the reciprocal of the network distance between $n_i$  and every other node in the network. That is, 
\begin{equation}
\label{equ:Closeness}
    C(n_i) = \sum_{j\in \mathcal{N}} \frac{1}{d(n_i,n_j)}
\end{equation}

\subsubsection{\label{subsubsec:Betweenness}Betweenness Centrality}
The betweenness centrality $B(n_i)$ measures how important a node $n_i$ is for traversing the network. It is calculated as:
\begin{equation}
\label{equ:Betweenness}
    B(n_i) = \sum_{\substack{n_j,n_k \in \mathcal{N}\\n_i\neq n_j \neq n_k}}\frac {\sigma_{n_j,n_k}(n_i)}{\sigma_{n_j,n_k}}
\end{equation}
where $\sigma_{n_j,n_k}$ is the number of shortest paths (i.e., the number of equivalent paths that have a smaller network distance than any other path) between nodes $n_j$ and $n_k$, and $\sigma_{n_j,n_k}(n_i)$ is the number of shortest paths that pass through $n_i$. In other words, the betweenness centrality is a calculation of the fraction of shortest paths between a pair of nodes that pass through $n_i$ summed over all pairs of nodes.

\subsubsection{\label{subsubsec:Eccentricity}Eccentricity}
The eccentricity $E(n_i)$ identifies the maximum distance between $n_i$ and every other node in the network. That is,
\begin{equation}
\label{equ:Eccentricity}
    E(n_i) = \max_{n_j \in \mathcal{N}}{d(n_i,n_j)}
\end{equation}

\subsubsection{\label{subsubsec:Katz}Katz Centrality}
The Katz centrality $K(n_i)$ is a measure that incorporates all of the paths in the system to determine the centrality of a node $n_i$. It is calculated in two equivalent ways, both using the weighted adjacency matrix:
\begin{equation}
\label{equ:Katz}
    K(n_i) = \alpha \sum_{n_j \in \mathcal{N}}A_{n_in_j}K(n_j) + 1 = \sum_{k=0}^\infty \sum_{n_j \in \mathcal{N}} \alpha^k (\textbf{A}^k)_{n_in_j}
\end{equation}
where the adjacency matrix $\textbf{A}$ is defined such that the element $A_{n_in_j} = 1$ if there is an edge between nodes $n_i$ and $n_j$ and otherwise $A_{n_in_j} = 0$. We use a the weighted adjacency matrix, which still sets $A_{n_in_j} = 0$ if there is no edge between nodes $n_i$ and $n_j$, but when an edge does exist between the nodes it sets $A_{n_in_j}$ to the weight of the edge connecting $n_i$ and $n_j$. $\alpha$ is an attenuation factor that must be less than the reciprocal of the largest eigenvalue of $\textbf{A}$ (i.e., $\alpha < 1/\lambda_{max}$). 

The Katz centrality can be interpreted as a measure of a node's importance by determining how connected it is to all of the other nodes in the network through every possible path. Longer paths are considered less important according to the value of $\alpha$. Thus, in this calculation, the importance of each path is dependent on both the weights on each edge in the path and on the number of edges in the path, with larger weights and smaller numbers of edges both being more important.

\subsubsection{\label{subsubsec:PageRank}PageRank}
The PageRank algorithm is an extension of the random walk that was originally designed by Larry Page to work as a search algorithm for Google \cite{Page1998, Bianchini2005}. It measures the likelihood that a random walker will end up at any given node. It differs from a standard random walk because at each step there is only a probability $\alpha$ that the walker traverses an edge and a probability $1-\alpha$ that the walker instead randomly jumps to any other node in the network. Mathematically, we can define a vector $\overrightarrow{\pi(t)}$ where each element $\pi_i(t)$ represents the probability that a random walker is at node $n_i$ at time step $t$. Thus, the PageRank algorithm is defined as:
\begin{equation}
\label{equ:PageRank}
\overrightarrow{\pi(t + 1)} = \alpha\textbf{P}\overrightarrow{\pi(t)} + (1-\alpha)\frac{\textbf{J}}{N}\overrightarrow{\pi(t)}
\end{equation}
where $\textbf{P}$ is a normalized weighted adjacency matrix that is left stochastic (i.e., each column is normalized to sum to 1) and $\textbf{J}$ is a $NxN$ matrix where every element is 1.

The values of the PageRank are calculated by finding the steady state of the PageRank algorithm $\overrightarrow{\pi*}$. That is, finding the vector such that:
\begin{equation}
\overrightarrow{\pi^*} = [\alpha\textbf{P}\ + (1-\alpha)\textbf{J}/N] \overrightarrow{\pi^*} 
\end{equation}
The values of this steady state vector are used as the centrality values of the nodes of the network.

\subsection{\label{subsec:PairCorrelation}Network pair correlation function}
In particle systems embedded in Euclidean space, the pair-correlation function (i.e., radial distribution function) $g_2(r)$ is a normalized measure of the density surrounding a given particle by measuring the probability that two particles are at a distance $r$ apart \cite{Torquato2018, jiao2010geometrical}. We can define the pair-correlation function mathematically in 2-D using the cumulative coordination number $Z(r)$ (i.e., the expected number of points within distance $r$ from a given point in the system) as:
\begin{equation}
\label{equ:pairCorr}
Z(r) = \rho 2 \pi \int_0^rxg_2(x)dx,
\end{equation}
where $\rho$ is the overall density of the system. The pair-correlation function indicates the probability of two particles to have a distance $r$ between their centers in comparison to an infinite Poisson distribution. In other words, an infinite Poisson distribution has a constant $g_2(r) = 1$, so a system that is more dense at distance $r$ has $g_2(r) > 1$ and a system less dense at distance $r$ has $g_2(r) < 1$. We translate this idea into network science by replacing the Euclidean distance between points with the distance based on network paths (i.e., $x$ in equation \ref{equ:pairCorr} is measured on the network instead of in Euclidean space). For the network pair-correlation function, we utilize the same normalization scheme as the point pattern pair-correlation function. That is, we still normalize $Z(r)$ by the overall density of the point pattern $\rho$ and the volume of a Euclidean ring $2\pi r dr$. Although this normalization scheme may not accurately capture the density evolution of the network, it can be used to compare $g_2(r)$ of the network to $g_2(r)$ of the point pattern.

\subsection{\label{subsubsec:DensityCorr}Local Density Covariance Function}
The local density covariance function is related to the pair-correlation function in that it provides a description of the density of the system, but instead of measuring the local density around a single point, it measures the correlation between the local densities of two points that are located at positions $\mathbf{r}_1$ and $\mathbf{r}_2$ with respect to the average local density. It can be described as:
\begin{equation}
C_{\rho_P}(\mathbf{r}_1,\mathbf{r}_2;\ell) = \frac{\langle \rho_P(\mathbf{r}_1;\ell)\rho_P(\mathbf{r}_2;\ell) \rangle-\langle \rho_P(\mathbf{r}_1;\ell) \rangle\langle \rho_P(\mathbf{r}_2;\ell) \rangle}{\sigma_{\rho_P(\mathbf{r}_1|\mathbf{r}_2;\ell)}\sigma_{\rho_P(\mathbf{r}_2|\mathbf{r}_1;\ell)}}
\end{equation}
Here, $\langle \cdot \rangle$ represents the ensemble average, and $\sigma_{\rho_P}$ represents the standard deviation of the distribution of local point pattern densities. Like the pair-correlation function, because our systems are isotropic, we can collapse this function to depend on the distance between the two points $C_{\rho_P}(r)$ for $r = |\mathbf{r_1}-\mathbf{r_2}|$. Qualitatively, this function allows us to observe if on average two points a distance $r$ apart have similar local densities, which informs how density fluctuates throughout the system. We also calculate a network version of this density function $C_{\rho_G}(r;\ell)$ where the local network density around every node $\rho_G(n_i;\ell)$ is used instead of the local point pattern density $\rho_P(\mathbf{r}_i;\ell)$ and the distance between points is determined based on the network distance $r = d(n_i,n_j)$.

\begin{figure*}[htpb]
    \includegraphics[width=\textwidth]{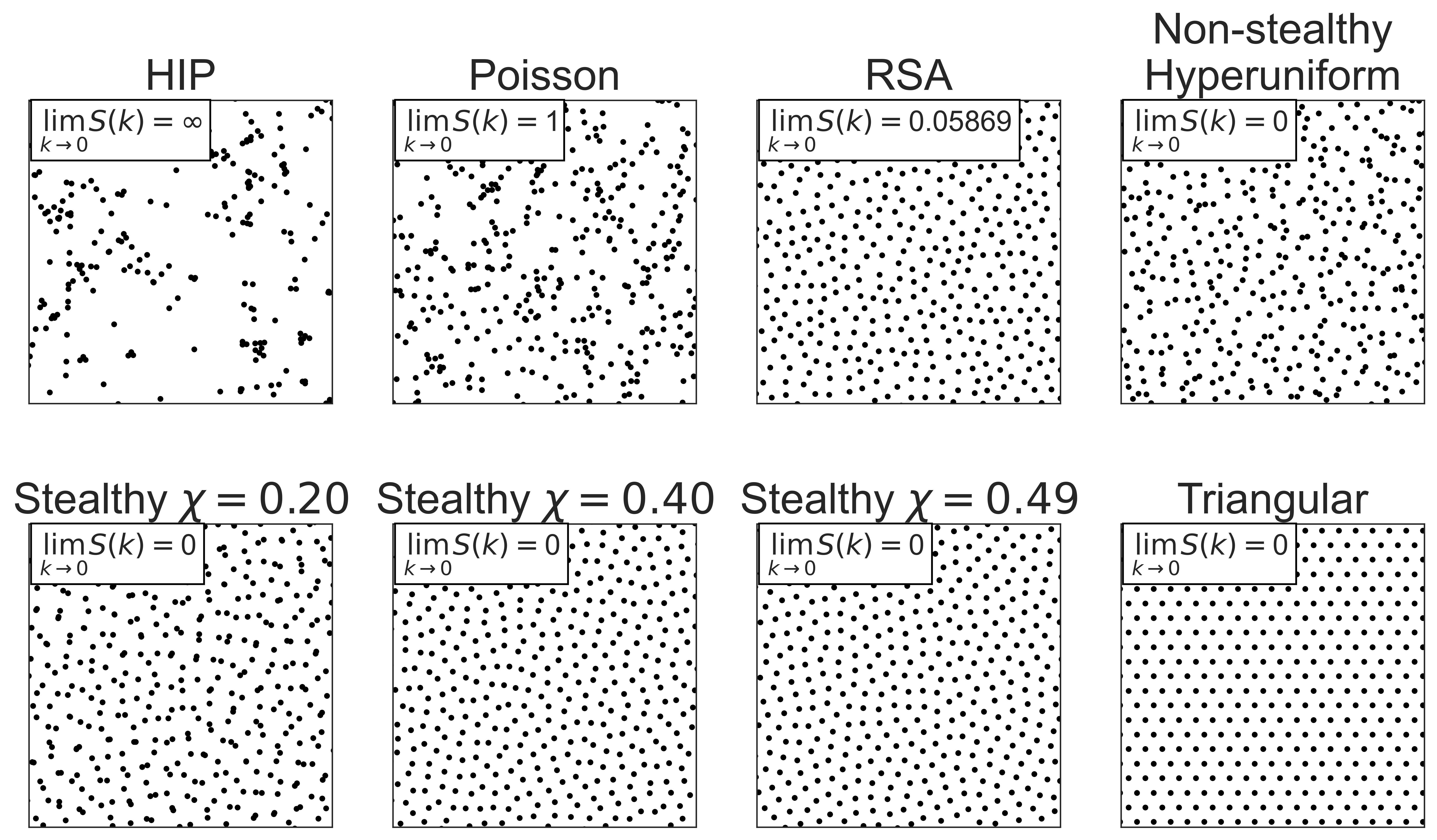}
    \caption{Representative plots of the model point configurations. The inset of each figure indicates the structure factor of the point configuration at the zero-wavenumber limit $\lim_{k\rightarrow 0}S(k)$.}
    \label{Fig:Represetative}
\end{figure*}

\section{\label{sec:Models}Models}
We determine the Delaunay triangulation of seven different classes of point patterns. We analyze one  antihyperuniform system obtained from the hyperplane intersection process (HIP) \cite{Stoyan1995}, two standard nonhyperuniform systems obtained respectively from Poisson and random sequential addition (RSA) \cite{Zhang2013,Torquato2006} processes, four disordered hyperuniform point patterns: one nonstealthy hyperuniform point pattern \cite{Wang2023, Wang2024} and three stealthy hyperuniform point patterns \cite{Torquato2015,Uche2004,Batten2008}. In applicable cases, we also use the triangular lattice as a benchmark. Figure \ref{Fig:Represetative} presents representative images of these eight point patterns. Each image is also labeled with the zero-wavenumber limit value of the structure factor $\lim_{k\rightarrow 0}S(k)$ for that point pattern, to indicate the different classes of systems.

The {\it hyperplane intersection process} (HIP) is the point process resulting from the intersection points of random, independent placements of hyperplanes. To generate this process in 2D, we find the intersections of a Poisson distribution of lines with random orientation and distance from the center of the unit cell. The resulting point process is hyperfluctuating. That is, the system is antihyperuniform because the number variance increases faster than the size of observation windows, which results in $lim_{k\rightarrow0}S(k) = \infty$.

A {\it Poisson point process} of $N$ points in 2D is generated by placing the $N$ points within a unit box with uniformly chosen points from $\mathbb{R}^2$. This results in a point pattern with structure factor $S(k) = 1$ for all $k$, which is nonhyperuniform. A {\it random sequential addition} (RSA) process in $d$-dimensions is a process of sphere packing in $\mathbb{R}^d$. Within the unit cell, hard $d$-dimensional spheres of fixed radius are sequentially added to the system in random positions, where the sphere is not placed if it would overlap with any previously placed spheres. Spheres are added until  the unit cell is maximally saturated. The structure factor of the 2D RSA packing has been numerically determined to be $\lim_{k\rightarrow 0}S(k) \approx 0.05869$. This value, while nearly zero, shows that the RSA packing is a nonhyperuniform system.

The {\it nonstealthy hyperuniform} point process is generated using a Gaussian pair statistic, with $g_2(r) = 1-exp(-\pi r^2)$. This model has $\alpha = 2$, which is a Class-I hyperuniform system, with $S(k) \sim k^2$ for small $k$ values. {\it Stealthy hyperuniform} systems are a special subset of Class-I hyperuniform systems possessing a zero structure factor for a range of wavevectors around the origin, i.e.,
\begin{equation}
S({\bf k}) = 0, \quad\text{for}\quad {\bf k} \in \Omega,
\label{eq_stealthy}
\end{equation}
excluding the forward scattering. Stealthy hyperuniform systems include all crystals and certain special disordered systems, which are characterized by a parameter $\chi$ reflecting the fraction of the constrained degrees of freedom in the system \cite{Morse2023, Torquato2018}, i.e.,
\begin{equation}
    \chi = N_\Omega/(N-d)
    \label{eq_chi_ratio}
\end{equation}
where $N_\Omega$ is the number of constrained degrees of freedom in the systems and $N$ is the total number of degrees of freedom. We note $d$ degrees of freedom associated with the trivial overall translation of the entire system are subtracted in Eq. (\ref{eq_chi_ratio}). In the case of point configurations, it has been shown that increasing $\chi$ leads to increased degree of order in the systems \cite{Morse2023, Torquato2018}. Disordered stealthy hyperuniform point configurations can be geneated using the so-called ``collective coordinates'' approach \cite{Morse2023, Torquato2018}.

For each of these seven types of systems, we generated a set of 10 different realizations for HIP, RSA and the stealthy hyperuniform systems, 6 different point patterns for the Poisson case, and 9 different point patterns for the nonstealthy hyperuniform case. Each point pattern consists of $\sim 10,000$ points within a unit square box. We subsequently generated the Delaunay triangulations of these point patterns using periodic boundary conditions (PBC). Because our point patterns are 2D and have a periodic boundary condition, every Delaunay triangulation has an average degree of 6, which follows from the Euler characteristic of a torus being zero. This property allows a comparison to the triangular lattice (i.e., an ordered hyperuniform system). Furthermore, to capture the information on the local neighborhood around each node, we characterize each edge of the networks with a weight indicating the Euclidean distance between the corresponding two points in the system.







\section{\label{sec:Results}Results}    
We analyze Delaunay triangulations built from seven different types of point patterns: HIP, Poisson, RSA, nonstealthy hyperuniform, and three stealthy hyperuniform. These systems are detailed in Section \ref{sec:Models}. We first determined the maximum tortuosity (ratio of the network distance and Euclidean distance) for each system. The highest tortuosity found across our systems was 1.46, the smallest maximum tortuosity of any system was 1.40, and the maximum tortuosities average to 1.42. Analysis of the maximum tortuosities indicates that these tortuosities result from network paths containing between 2 and 4 edges. That is, the largest difference between the network distances and the Euclidean distances occurs for short paths, which indicates that the high tortuosity is due to the short-range jaggedness of these triangular networks. At longer path lengths the tortuosity decreases because at longer length-scales this local noise is smoothed out by the increased length of the path. Thus, our systems do not reach tortuosities near the theoretical maximum (i.e., between 1.59 and 1.84), which indicates that these systems are structurally different from the typically studied Delaunay triangulation extremes. Despite this distinction, the networks do distort the underlying Euclidean distances of the point pattern.

\subsection{\label{subsec:CentResults}Centrality Results}
On these networks we investigate a variety of network metrics (see Key Concepts) to describe the degree of order in the Delaunay triangulations. Specifically, we interpret the dispersion of the centrality metric’s distribution as a descriptor of the network’s position on a spectrum between order (i.e., low dispersion) and disorder (i.e., high dispersion).  We test a variety of centrality measures to investigate how successful various measures are at capturing the order or disorder of each system. These centrality measures are chosen to have varying length scales to attempt to capture the various length scales of order present in our systems. We specifically investigate weighted centrality measures because the inclusion of edge weights assists in capturing the local neighborhoods of the original point pattern. The dispersions of each system's triangulations are compared to each other to determine if a given centrality metric can recapitulate the known placement of the underlying point patterns on the spectrum between order and disorder. The extent of success of each centrality metric also provides insight into what aspects of the point pattern are reproduced by the triangulation.

\begin{figure*}[htpb]
    \includegraphics[width=\textwidth]{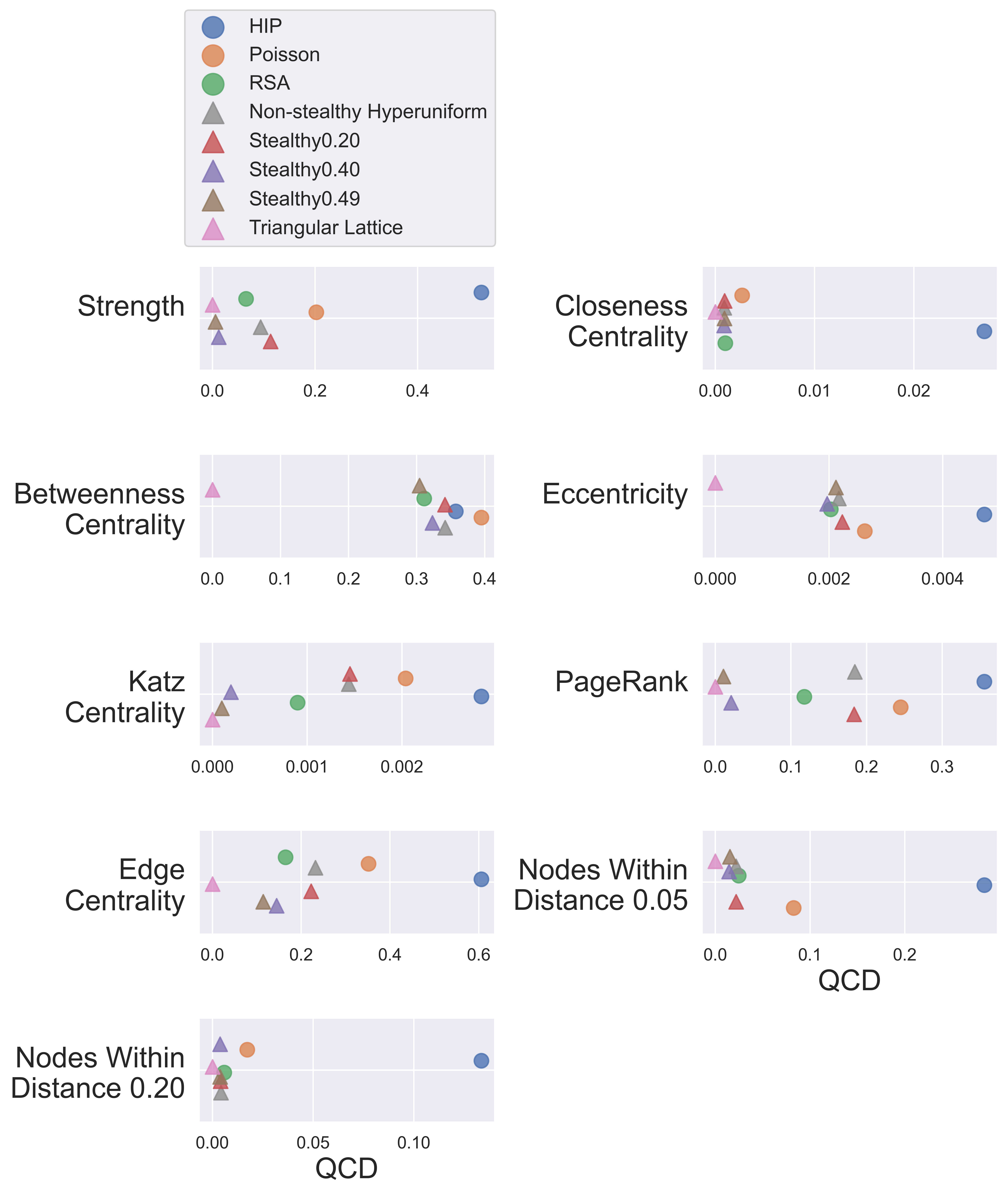}
    \caption{Quartile coefficient of dispersion (QCD) for the distributions of various centrality measures values. QCDs are calculated for each of the seven network types, and we include the triangular lattice as a point of comparison. Hyperuniform networks are represented with a triangular symbol, and nonhyperuniform networks are represented with a circular symbol. We note that because each metric has a different algorithm, the QCD values are not comparable between the different metrics. We also not that the y values of these points have no inherent meaning and are only used to visually separate the points.}
    \label{Fig:Cent}
\end{figure*}

For each centrality metric, we calculate every system’s quartile coefficient of dispersion (QCD $= (Q_3-Q_1)/(Q_3+Q_1)$) based on the distributions of centrality values of every node in every network of a given type. The QCD is a robust measure of scale that is used to compare the amount of dispersion between distributions. To act as a point of comparison, we calculate the QCD of each centrality measure on the triangular lattice (i.e., an ordered hyperuniform system). Figure \ref{Fig:Cent} presents the QCDs for every centrality measure. Hyperuniform systems are represented with triangular symbols, and the nonhyperuniform systems are represented as circles. We note that each of the centrality metrics are calculated using a different algorithm, so the values of the QCDs of different metrics should not be compared.

Because the triangular lattice is a regular network, it receives a QCD of 0 for every centrality measure (i.e., there is no dispersion because every node is equivalent). Among the centrality measures, all, except the betweenness centrality, correctly place the HIP triangulation as the most heterogeneous of the systems and the Poisson triangulation as the second most heterogeneous. The reason for this discrepancy in the betweenness centrality is likely the fact that the betweenness centrality depends on a network sharing many equal-length paths, which does not occur as often when calculating weighted paths. Indeed, the HIP does have the highest QCD for the unweighted betweenness centrality. Many of the centrality measures — strength, betweenness centrality, Katz centrality, PageRank, and edge centrality — place the three stealthy hyperuniform systems in order, with the stealthy hyperuniform systems with lower $\chi$ values having higher QCDs. This ranking recapitulates the analogous result for point patterns \cite{Torquato2003}. These centrality measures all measure short to moderate length-scales, and three of them — betweenness centrality, Katz centrality, and PageRank — involve measuring the variety in paths through the systems. In the cases where the three stealthy hyperuniform systems are incorrectly placed, the deficiency is in distinguishing between the $\chi = 0.40$ and $\chi = 0.49$ systems. We conclude that the placements of the QCDs of some centrality measures can provide a means for rank-ordering the hyperuniformity of these various triangulations, but this is not universally true for all centrality measures.

The centrality metrics do not fully capture some of the more subtle differences between the systems. Specifically, the RSA system, while nonhyperuniform, typically has a QCD that lies in between the stealthy $\chi = 0.20$ and stealthy $\chi = 0.40$ systems. The RSA point pattern is considered a nearly-hyperuniform system (i.e., the RSA’s normalized variance becomes very small in the infinite limit, but it approaches the constant 0.0589 instead of zero \cite{Atkinson2016,Torquato2021PRE}), so this placement near the most disordered hyperuniform systems is not surprising. The fact that the RSA is placed in this location for almost all the centrality metrics indicates that the centrality metrics are incapable of fully differentiating between hyperuniform and nonhyperuniform systems. That is, for a system that is very similar to, but has subtle differences from hyperuniform systems, we get a triangulation that is statistically identical, based on the centrality values, to the triangulation of hyperuniform systems. This result, along with the errors in the placements of hyperuniform systems for some centrality metrics, suggests that the Delaunay triangulation may not sufficiently capture density information of the underlying point pattern, leading to a loss of the subtle differences between the properties of the hyperuniform point patterns and of the RSA. Indeed, our comparisons between the network density and density of the underlying point pattern, which we describe in the next section, demonstrate that the Delaunay triangulation distorts density in hyperuniform point patterns to reduce the suppression of density fluctuations and inversely distorts the density in the RSA systems to increase the suppression of density fluctuations. These distortions render the hyperuniform triangulations more similar to the RSA system’s triangulation.

\subsection{\label{subsec:DensityRes}Density Comparison}
\begin{figure*}[htpb]
    \includegraphics[width=\textwidth]{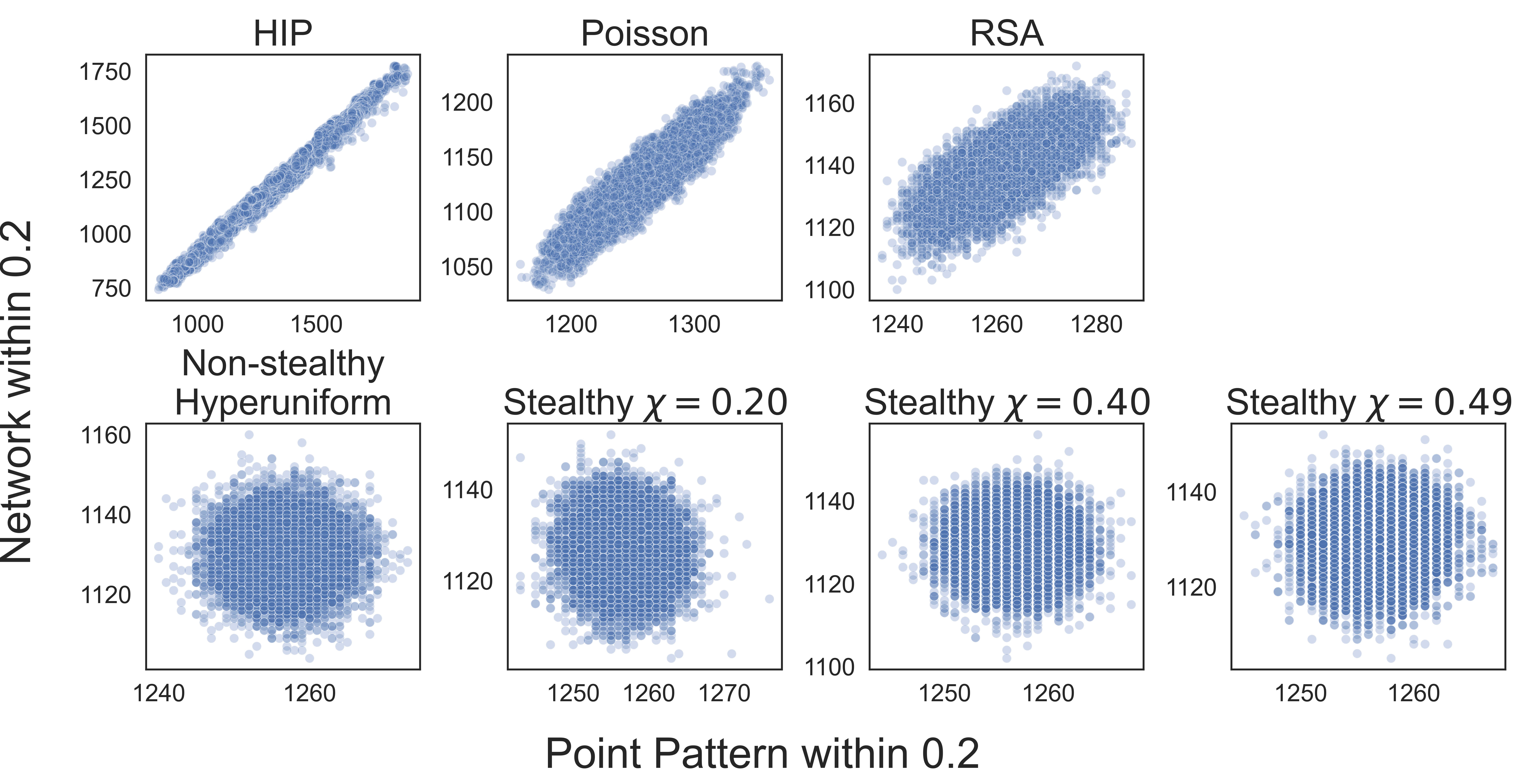}
    \caption{Density comparison between the Delaunay triangulation and its underlying point pattern. For each point in a system, we find the number of points within a Euclidean distance of 0.20 in the point pattern (x-axis) and plot it against the number of nodes within a network distance of 0.20 of the corresponding node in the Delaunay triangulation (y-axis). The network distance is calculated only along the weighted edges of the network, which is why it does not have the same density values as the point pattern. The nonhyperuniform systems are in the top row, and the hyperuniform systems are in the bottom row.}
    \label{Fig:DensityScatter}
\end{figure*}
\begin{figure*}[htpb]
    \includegraphics[width=\textwidth]{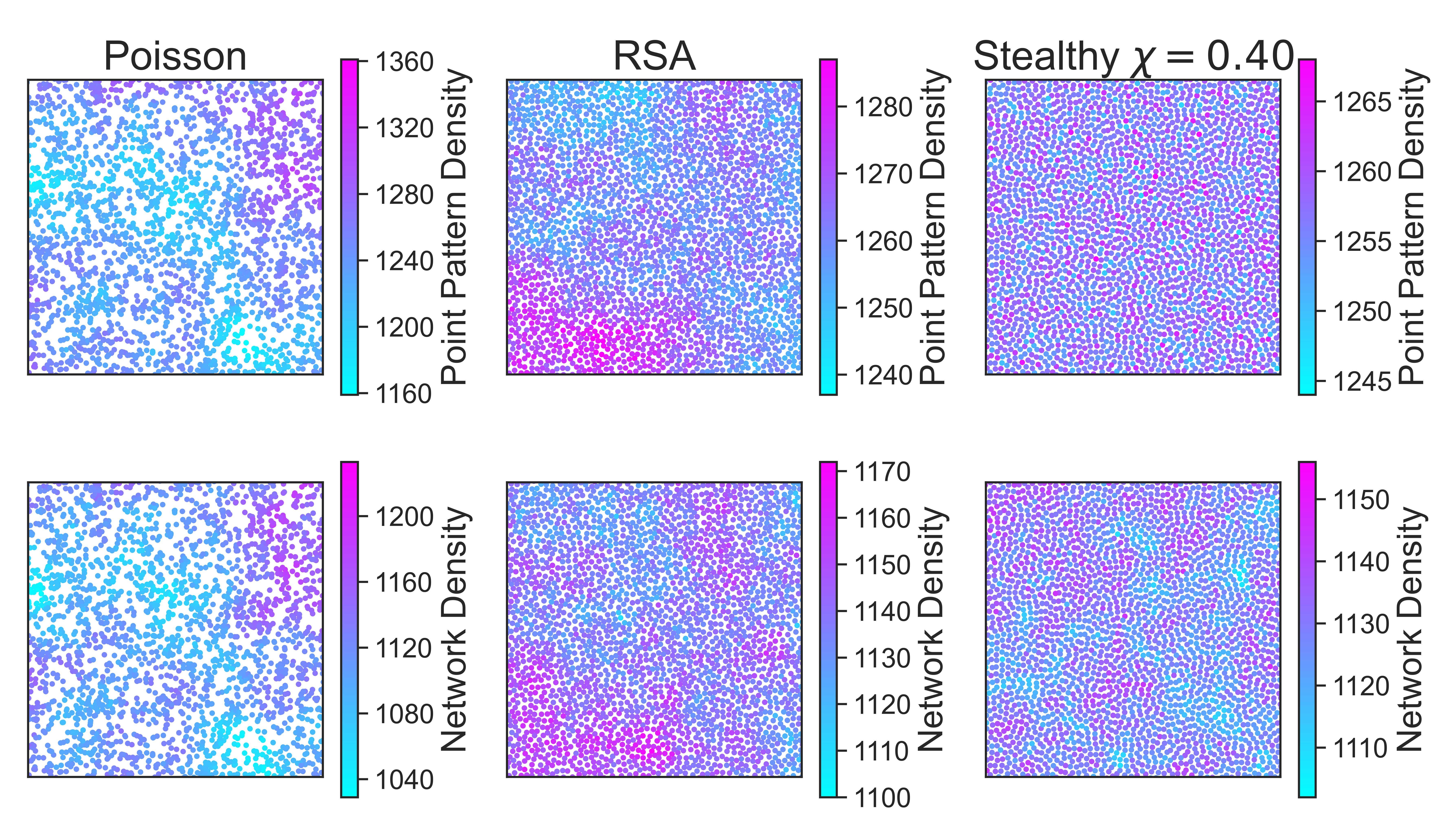}
    \caption{Illustrative examples of the point pattern density (top row) and network density (bottom row) of the Poisson, RSA, and stealthy $\chi = 0.40$ systems. In each plot, the local density within a distance of 0.05 (i.e. $\rho_P(\mathbf{r}_i;0.05)$ and $\rho_G(n_i;0.05)$ for the point pattern and network respectively) is indicated on a color scale, with purple representing the highest local density and blue representing the lowest local density. We note the high degree of similarity in the coloring of the point pattern and network plots for the nonhyperuniform systems, which is not present in the hyperuniform system.}
    \label{Fig:DensityComparison}
\end{figure*}

To investigate the degree of  the Delaunay triangulation’s reproduction of the density fluctuations of the system, we compare the local Euclidean density $\rho_P(\mathbf{r}_i;\ell)$ around each point $\mathbf{r}_i$ to the local network density $\rho_G(n_i;\ell)$ around the corresponding node $n_i$. Figure \ref{Fig:DensityScatter} showcases these comparisons as scatterplots of $\rho_G(n_i;\ell)$ vs $\rho_P(\mathbf{r}_i;\ell)$ for $\ell = 0.20$ for every point/node in each system. From Figure \ref{Fig:DensityScatter}, it is clear that the antihyperuniform and nonhyperuniform systems have a correlation between the point pattern density and network density, but no correlation is  evident for the hyperuniform systems. We provide a visual of this difference of densities in Fig. \ref{Fig:DensityComparison}, which shows the local densities of the point patterns on the top and the local network densities on the bottom for the Poisson system, RSA system, and the stealthy $\chi = 0.40$ system. The correlation is visually apparent in the Poisson and RSA systems, with the dense sections of the point pattern also being dense in the triangulation, but this is not true for the hyperuniform system, where there is no relationship between the densities in the two representations.

\begin{figure*}[htpb]
    \includegraphics[width=\textwidth]{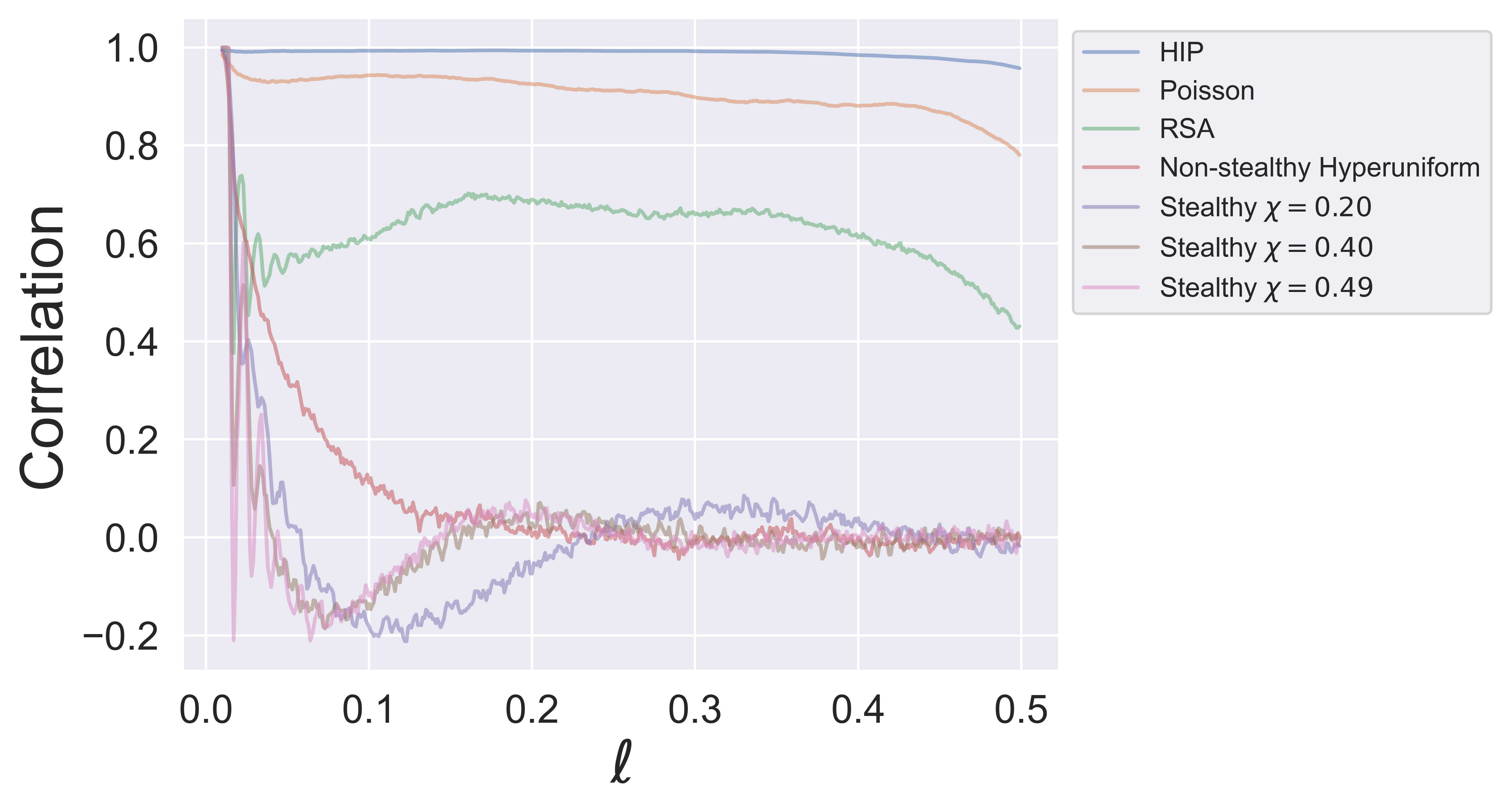}
    \caption{Pearson correlation coefficients (PCC) of the network and point pattern densities for increasing distances $\ell$. For each system, we calculate the PCC between the network densities (i.e., the number of nodes with a network distance $\leq \ell$ for each node) and the point pattern densities (i.e., the number of points with a Euclidean distance $\leq \ell$ for each point). We calculate these PCC values for $\ell$ values between 0 and 0.5 because for $\ell > 0.5$ we begin to see finite size effects.}
    \label{Fig:Correlation}
\end{figure*}

We note that the hyperuniform systems cover a smaller range of values in both densities than the nonhyperuniform systems. The limited range of both densities may contribute to, but does not explain, the lack of correlation. To assess if the lack of correlation in hyperuniform systems is a result of this smaller range of values, we calculate the Pearson correlation coefficient (PCC) between the densities of the triangulations $\rho_G(n_i)$ and of the point patterns $\rho_P(\mathbf{r}_i)$ at various selection radii $\ell$. Figure \ref{Fig:Correlation} shows the evolution of the PCC for each system type over $\ell$ values from 0 to 0.5. We do not consider $\ell$ values greater than 0.5 because radii larger than 0.5 begin to see the edge effects of these finite systems. It is clear from Figure \ref{Fig:Correlation} that the trend in the PCCs for hyperuniform systems is qualitatively different from the trend of nonhyperuniform systems. Every system begins with a PCC of 1 because the triangulation's weighted edges preserves the local information in the nearest neighbors of the point pattern. The PCC of  the HIP, Poisson and RSA systems remains high over the whole range of $\ell$. In contrast, the PCC of hyperuniform systems quickly decreases to 0, indicating that the triangulation’s density is not correlated to its underlying point pattern beyond the local level. This result indicates that the weighted Delaunay triangulation does not accurately capture the density of hyperuniform point patterns beyond the local length scale.

We also note that the correlations of the stealthy hyperuniform systems in Fig. \ref{Fig:Correlation} show damped oscillations around the correlation of zero. This feature indicates that at certain $\ell$ values the stealthy hyperuniform Delaunay triangulation is worse at capturing the denser parts of the point pattern, causing the two densities $\rho_P(\mathbf{r}_i)$ and $\rho_G(n_i)$ to be anticorrelated. At larger $\ell$ values this anticorrelation becomes a slight correlation because the system is connected, causing the newly included nodes to be anticorrelated with the original anticorrelated nodes. Because of the long-range order of these systems, as $\ell$ increases the local densities $\rho_P(\mathbf{r}_i)$ and $\rho_G(n_i)$ become more uniform, which causes the oscillations to dampen. This result indicates that further investigation of the density of the network $\rho_G$ in comparison to the underlying point pattern $\rho_P$ is needed to reveal how the structure of the stealthy triangulations affect their ability to capture the density of the original point pattern.

\subsection{\label{subsec:NetworkPairResults}Network pair correlation function}
\begin{figure*}[htpb]
    \includegraphics[width=\textwidth]{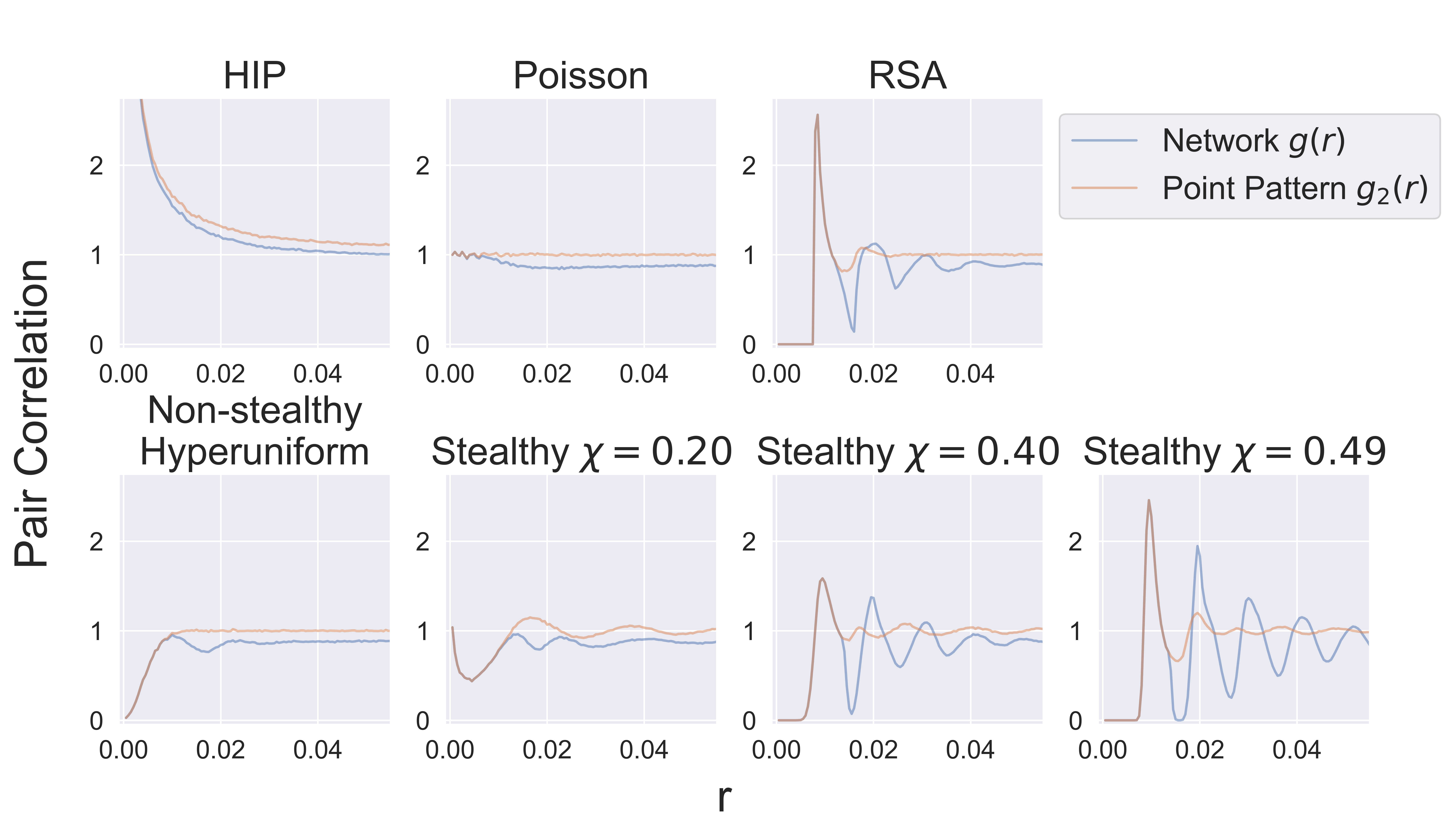}
    \caption{Network pair-correlation function for each system. The pair-correlation function provides the conditional probability of finding a point at distance $r$ from a given point at the center. We extend this concept to networks by finding the probability that a node will be at network distance $r$ from another node. To normalize the network pair-correlation function, we use the same normalization as the Euclidean pair-correlation function; this normalization causes the functions to approach values slightly less than 1.}
    \label{Fig:PairCorr}
\end{figure*}

To determine why the Delaunay triangulation is unsuccessful at reproducing the density of hyperuniform point patterns, we investigate a variety of correlations on our various systems. The first function that we investigate is the network pair-correlation function. This function reveals the general density distribution around nodes within the network. The triangulation, by nature, stretches the distance between points in comparison to the point pattern, so we expect that the network pair-correlation function $g(r)$ can be used to understand how the triangulations distort density when compared to the point pattern pair-correlation function $g_2(r)$. Figure \ref{Fig:PairCorr} shows the network pair-correlation functions (blue) and point pattern pair-correlation functions (orange) for each system. From these plots, we can see that initially the network function is equivalent to the point pattern function, but these curves begin to differentiate as $r$ increases. Over all $r$, every network pair correlation function maintains the same form as its corresponding point pattern function. The antihyperuniform HIP system’s pair correlation decreases from infinity as $r$ increases, the Poisson system’s is nearly constant for any $r$, the nonstealthy hyperuniform system’s pair correlation increases to a constant as $r$ increases, and the RSA and stealthy hyperuniform systems showcase damped oscillations. The damped oscillations are a result of the more binned nature of the distances between points. Furthermore, traversing the network on edges of similar lengths more rigidly bins the distances causing the oscillations in the network pair correlation function to be larger.

Despite the network correlation functions having the same functional form, there are differences that result from the network structure. The network pair-correlation function ultimately decreases below the point pattern pair-correlation function. This happens because the network pair-correlation function is normalized using the same ring area as the original pair-correlation function. The stretching of the distances in the triangulation effectively reduces the measured density of the network. This causes the network pair-correlation function to approach values slightly lower than 1 at large $r$, which is the limit of the point pattern pair-correlation functions. When analyzing the limits of the network functions as $r$ approaches infinity (Table I), we see that the stealthy hyperuniform systems have the lowest limits and the antihyperuniform HIP system has the highest limit. This indicates that the hyperuniform systems have a slightly higher distortion between the Euclidean distance and the network distance between points. We note that the differences between these limits are not significant enough to indicate large structural differences between the different networks. Nevertheless, this slight difference is expected because a lattice (which the stealthy hyperuniform systems resemble) constrains the directions that can be traversed, which causes more tortuous paths (e.g., in a square lattice all edges are horizontal or vertical, but pairs of nodes separated by both horizontal and vertical edges have a much shorter Euclidean distance on the inaccessible diagonal between the two nodes). This increased stretching in the hyperuniform systems is indicative that these systems create larger distortions of the underlying density in the construction of the triangulation, which, in turn, causes the density fluctuations in the hyperuniform systems to be masked.

\begin{table*}[!t]
\setlength\tabcolsep{1.pt}
\caption{Limits of network pair-correlation functions as $r$ approaches infinity for each Delaunay triangulation.}
{\begin{tabular*}{\textwidth}{@{\extracolsep{\fill}}lccccccc}
\hline
Network&HIP & Poisson & RSA & \shortstack{nonStealthy\\Hyperuniform}&\shortstack{Stealthy\\$\chi=0.20$} & \shortstack{Stealthy\\$\chi=0.40$} & \shortstack{Stealthy\\$\chi=0.49$}\\
\hline
\shortstack{$\lim_{r\rightarrow\infty}g(r)$} & 0.9443 & 0.9134 & 0.9173 & 0.9139 & 0.9120 & 0.9131 & 0.9125\\
\end{tabular*}}{}
\end{table*}

\subsection{\label{subsec:DensityCorrRes}Local Density Covariance Function}
\begin{figure*}[htpb]
    \includegraphics[width=\textwidth]{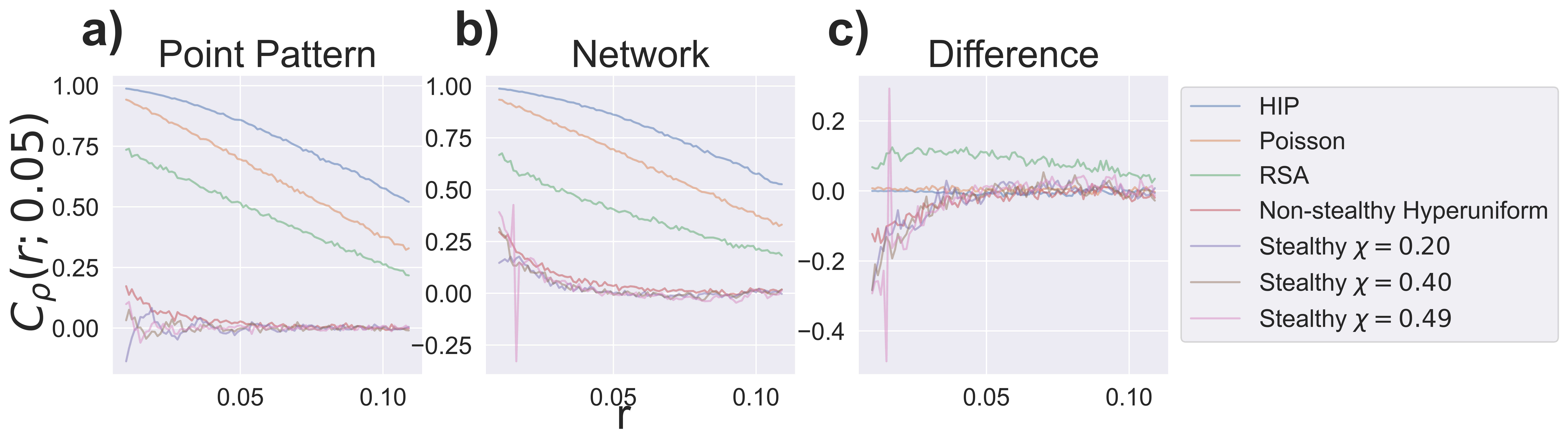}
    \caption{Local density covariance function for the point pattern $C_{\rho_P}(r;0.10)$ (a) and the network $C_{\rho_G}(r;0.10)$ (b). We also calculate the difference between the point pattern local density covariance function and network local density covariance function $C_{\rho_P}(r;0.10)-C_{\rho_G}(r;0.10)$ (c). The local density covariance function quantifies the correlation of local densities $\rho_P(\mathbf{r}_i)$ ($\rho_G(n_i)$) between two points (nodes) at a distance $r$ apart. Here we calculate the local density using $\ell = 0.10$, which is large enough to avoid the initial noise of small distances while also still being local to each point (node).}
    \label{Fig:DensityCovariance}
\end{figure*}

Calculating local density covariance $C_\rho(r;\ell)$ functions allows us to compare the density fluctuations of the point pattern with those of the network. Figure \ref{Fig:DensityCovariance} plots each system’s local density covariance function for the point pattern $C_{\rho_P}(r;0.10)$ (panel a), local density covariance function for the network $C_{\rho_G}(r;0.10)$ (panel b), and the difference between the point pattern local density covariance function and network local density covariance function $C_{\rho_P}(r;0.10)-C_{\rho_G}(r;0.10)$ (panel c). For these plots, we choose a $\ell$ value of 0.10, to be large enough to capture the surrounding density of each point without having too much noise and small enough that the number variance of the hyperuniform systems is not too small. The point pattern and network local density covariance functions both showcase a qualitative difference between the hyperuniform systems and the nonhyperuniform systems: the hyperuniform systems have much lower covariance values across all $r$. These near zero covariance values at almost all distances indicate the suppression of density fluctuations in the hyperuniform systems. For the HIP, Poisson, and RSA systems, the local density covariance is positive and decreasing with distance in both the point pattern $C_{\rho_P}(r)$ and the network $C_{\rho_G}(r)$. For the HIP and Poisson systems, these functions are are very similar with a small difference between the network and point pattern functions. While still similar overall, the RSA system is less similar, and it is the only system in which the local density covariance is smaller in the network than in the point pattern. These positive difference values in Fig. \ref{Fig:DensityCovariance}c indicate that the triangulation suppresses density fluctuations more than the point pattern. The local density covariance functions for the hyperuniform systems have a different shape when comparing the point pattern covariance function $C_{\rho_P}(r)$ and the network covariance function $C_{\rho_G}(r)$. For the hyperuniform systems the point pattern local density covariance function $C_{\rho_P}(r)$ has very small values across all $r$ while the network's local density covariance function $C_{\rho_G}(r)$ begins with positive correlations that decrease to zero as $r$ increases. The negative values of the differences for the hyperuniform systems indicate that the triangulation is worse at suppressing density fluctuations than the point pattern. Thus, in the construction of the Delaunay triangulation an added covariance at small $r$ is manufactured.

This result elucidates why the Delaunay triangulation is ineffective at capturing the density fluctuations of hyperuniform systems. Because disordered hyperuniform point patterns are characterized by suppression of normalized large-scale density fluctuations while also being statistically isotropic with no Bragg peaks, the local densities of neighboring points in the point pattern are not correlated. That is, the immediate neighbohood (i.e., small $r$) of each point can have very different local density values $\rho_P(\mathbf{r}_i)$ in the point pattern. This property can be seen in Fig. \ref{Fig:DensityComparison}, where the stealthy $\chi = 0.40$ system has points of opposite density neighboring each other. On the other hand, in the triangulation, each node is connected to its immediate neighbors, so the local network density $\rho_G(n_i)$ of one node cannot be substantially different from its neighbor, which manifests as the positive local density covariance for small $r$ in the networks. The rapid decay of the network covariance functions for hyperuniform systems indicates that the triangulation effectively suppresses large $r$ density fluctuations. In summary, the rapid density fluctuations between neighbors of the point pattern cannot be replicated in the Delaunay triangulation by nature of the construction of the network, which results in the density of the triangulation not mirroring the density of the point pattern. Conversely, because the other, nonhyperuniform point patterns have positive covariance at small $r$, the network is capable of reproducing the density fluctuations of these systems.

\section{\label{sec:Discussion}Discussion}
In this manuscript, we have investigated the Delaunay triangulation as a means of translating point patterns into spatial networks. Because the distance metric is changed in the translation of the point pattern into a network, we specifically investigated how effective the Delaunay triangulation is at capturing the local densities $\rho_P(\mathbf{r}_i)$ of the point patterns by analyzing the correlations between $\rho_P(\mathbf{r}_i)$ and the local network density $\rho_G(n_i)$ and how effective the triangulation is at reproducing the fluctuations in local densities between points by analyzing the local density covariance of the point pattern $C_{\rho_P}(r)$ and the network $C_{\rho_G}(r)$. We studied seven different types of point patterns, which can be sorted into three classes, antihyperuniform, nonhyperuniform, and hyperuniform. We were interested in better understanding whether, and how the Delaunay triangulation distorts the local densities of the point pattern. Understanding these distortions determines the efficacy of using the Delaunay triangulation as a spatial network tool to investigate embedded systems. 

We were motivated to do this investigation because our analysis of the heterogeneity of centrality measures was not able to identify a centrality measure that recapitulated the results of a number variance analysis of the point patterns. Some centrality measures were unable to accurately sort hyperuniform networks into the expected sequence, and none of the centrality measures were able to completely differentiate hyperuniform triangulations from nonhyperuniform triangulations. Specifically, the nonhyperuniform RSA system achieved QCD values similar to the stealthy hyperuniform systems in every centrality measure. Thus, while some centrality measures were effective at rank-ordering the various hyperuniform and nonhyperuniform systems within their own class (e.g., the Katz centrality and PageRank), they were all ineffective at separating the QCD values of the RSA system from the hyperuniform systems. Since we tested a diverse population of centrality metrics with a variety of length scales, we believe that the inability of any of these metrics to classify the systems indicates that there is a systematic error of the Delaunay triangulation. This error prevents the triangulation from accurately capturing the subtle density features of the hyperuniform systems.

We applied a variety of measures to our systems to elucidate how the local densities of each point are distorted in their respective networks. We first investigated how the local density of each point $\mathbf{r}_i$ in the point pattern $\rho_P(\mathbf{r}_i)$ correlates with the local density of the corresponding node $n_i$ in the network $\rho_G(n_i)$. This initial analysis confirmed that the Delaunay triangulation does not accurately capture the density of the hyperuniform systems. As shown in Fig. \ref{Fig:Correlation}, when comparing densities at small $\ell$ there is a correlation between the network density and the density in the point pattern because the Delaunay triangulation captures local information in the edge weights of the system, but as $\ell$ increases, the hyperuniform systems’ correlations rapidly decrease to a nearly uncorrelated state. The stealthy hyperuniform systems also display interesting oscillations around the correlation value of 0, systems with higher $\chi$ values having a higher frequency. The determinants of this frequency and their relationship to the Delaunay triangulation in stealthy systems provides a future avenue to explore. Overall, these correlation results indicate that there is a stark difference in the ability of the Delaunay triangulation to capture the density of point patterns, with a high fidelity for nonhyperuniform systems and a low fidelity (i.e., no correlation at large $\ell$) for hyperuniform systems.

To further explore the density of these systems, we utilized two different functions. First, the network pair-correlation function allowed us to probe how the overall density of the system is distorted by the Delaunay triangulation, and second, the local density covariance function allowed us to probe how the density’s fluctuations are distorted by the Delaunay triangulation. The network pair-correlation function indicated that the hyperuniform systems are distorted slightly more than the nonhyperuniform systems, with this increased distortion likely contributing to the loss of density information of the underlying point pattern. Along with this increased distortion, comparison of the local density covariance functions $C_\rho(r)$ indicated that in the hyperuniform networks, two nodes that are close together (i.e., small $r$) have correlated local densities (i.e., $C_{\rho_G}(r) > 0$), while their corresponding points’ local densities are not correlated in the point pattern (i.e., $C_{\rho_P}(r) \approx 0$). This indicates that a major contributor to the density distortion is the fact that the network is a connected entity, in contrast to the independent points of the point pattern, so the neighbors of a node cannot have drastically different densities.

Through this density analysis we show that the construction of the Delaunay triangulation is ineffective at capturing the local densities and density fluctuations of disordered hyperuniform systems, but it is still effective at capturing these properties in other systems. Disordered hyperuniform systems are similar to liquids or glasses in that they are statistically isotropic and possess no Bragg peaks, and yet they completely suppress normalized large-scale density fluctuations, like crystals, and in this sense, possess a hidden long-range order. As evidenced by our results, the Delaunay triangulation cannot replicate this unique structural property of hyperuniform systems because the network connects neighboring points. This causes the disordered hyperuniform systems to have local network densities that are correlated, which differs from the point patterns that have no local point pattern density correlations at small $r$. In contrast, nonhyperuniform systems have point patterns that have correlations at small $r$ (i.e., $C_{\rho_P}(r) > 0$ for small $r$). Thus, the edges of the Delaunay triangulation connecting points that have similar local point pattern densities $\rho_P(\mathbf{r}_i)$ results in nodes that have similar local network densities $\rho_G(n_i)$, which makes the Delaunay triangulation of nonhyperuniform systems capable of reproducing the density fluctuations of the underlying point pattern. 

\begin{table*}[!t]
\setlength\tabcolsep{1.pt}
\caption{Euclidean density assortativity of the Delaunay triangulations.}
{\begin{tabular*}{\textwidth}{@{\extracolsep{\fill}}lccccccc}
\hline
Network&HIP & Poisson & RSA & \shortstack{nonStealthy\\Hyperuniform}&\shortstack{Stealthy\\$\chi=0.20$} & \shortstack{Stealthy\\$\chi=0.40$} & \shortstack{Stealthy\\$\chi=0.49$}\\
\hline
\shortstack{Euclidean\\Density\\Assortativity} & 0.993 & 0.958 & 0.823 & 0.179 & -0.0277 & 0.0167 & 0.0559\\
\end{tabular*}}{}
\end{table*}

While this result shows that information derived from a Delaunay triangulation is not sufficient to determine if a system is hyperuniform, there is still a rich potential for the intersection of network methods with hyperuniformity. Current work is being done to use network information in conjunction with the point pattern to study hyperuniformity \cite{Newby2024}. Networks can provide alternative ways to study hyperuniformity, resulting in a better understanding of the unique properties of these systems. For instance, our centrality analysis did showcase that path-based metrics were more successful than other centrality measures, which indicates that hyperuniformity is likely related to transport through the network. We expect further analysis of the paths in  other spatial networks built from hyperuniform systems, for example geometric networks, will reveal properties of efficient transport in hyperuniform systems, which is a question that is currently being studied in point patterns and two-phase materials \cite{Cheron2022, Shi2023,Kim2024,Liang2024,Garcia2021}. Analysis of the triangulation’s Euclidean density assortativity (i.e., the Pearson correlation coefficient of the local point pattern densities $\rho_P(\mathbf{r}_i)$ between points attached by an edge), shown in Table II, indicates that the disordered hyperuniform networks do not showcase Euclidean density correlations with their direct neighbors, which contrasts with the nonhyperuniform systems. This property, which is only calculable on the network, is useful in better understanding how each point in a disordered hyperuniform system interacts with its neighbors. Further study of similar properties will yield a better understanding of these systems. An especially relevant topic of investigation is understanding the exclusionary zones that are present in multi-hyperuniform systems \cite{Jiao2014}. Thus, we believe that while network measures may not be useful classifiers, the novel study of disordered hyperuniform systems using network-based analyses is an interesting prospect with many open questions that require further investigation and the potential to better understand this unique state of matter.

We conclude that the Delaunay triangulation is an effective model of the underlying point pattern only when the local densities $\rho_P(\mathbf{r}_i)$ of points that are located close together in the point pattern are correlated (i.e., $C_{\rho_P}(r) > 0$ for small $r$). Fortunately, a point pattern having correlated local densities is a general property, and the exotic state of matter of disordered hyperuniform systems is the first discovered and currently the only system that is an exception from it. It is of future interest how this approach performs for other spatial network algorithms, such as geometric networks, to further determine the applicability of spatial networks in density-based analyses. Overall, the Delaunay triangulation is a useful tool for many real-world applications and it is appropriate for spatial network analysis in most systems because it will reproduce the underlying density of the point pattern.

\section*{Acknowledgements}
This work was supported by the Army Research Office under Cooperative Agreement Number W911NF-22-2-0103.

\bibliographystyle{unsrtnat}
\bibliography{Del_references}

\end{document}